\documentclass[12pt]{article}
\usepackage{epsfig}

\parskip=\medskipamount 
plus.3em minus.5em \abovedisplayshortskip=.5em plus.2em minus.4em
\renewcommand{\thesection}{\arabic{section}}

\catcode`@=11

\@addtoreset{equation}{section} \@addtoreset{equation}{subsection}
\def\theequation{\ifnum\value{section}=0 \arabic{equation}\ignorespaces
\else \ifnum\value{section}=-1 A.\arabic{equation}\ignorespaces
\else \ifnum\value{subsection}=0
\thesection.\arabic{equation}\ignorespaces \else
\thesection.\arabic{subsection}.\arabic{equation}\ignorespaces
                             \fi
                        \fi
                   \fi}

{\catcode`\'=\active \def'{{}^\bgroup\prim@s}}

\catcode`@=12

\newcommand{\bq}{\begin{equation}}
\newcommand{\be}{\begin{equation}}
\newcommand{\fq}{\end{equation}}
\newcommand{\ee}{\end{equation}}
\newcommand{\bqr}{\begin{eqnarray}}
\newcommand{\beqs}{\begin{eqnarray}}
\newcommand{\fqr}{\end{eqnarray}}
\newcommand{\eeqs}{\end{eqnarray}}

\newcommand{\rf}[1]{(\ref{#1})}

\def\bop#1{\setbox0=\hbox{$#1M$}\mkern1.5mu
    \vbox{\hrule height0pt depth.04\ht0
    \hbox{\vrule width.04\ht0 height.9\ht0 \kern.9\ht0
    \vrule width.04\ht0}\hrule height.04\ht0}\mkern1.5mu}

\begin{document}
\thispagestyle{empty}

\begin{flushright}
\begin{tabular}{l}
hep-th/0507207 \\
\end{tabular}
\end{flushright}

\vskip .6in
\begin{center}

{\bf A Count of Classical Field Theory Graphs}

\vskip .6in

{\bf Gordon Chalmers}
\\[5mm]

{e-mail: gordon@quartz.shango.com}

\vskip .5in minus .2in

{\bf Abstract}

\end{center}

A generating function is derived that counts the number of diagrams in an 
arbitrary scalar field theory.  The number of graphs containing 
any number 
$n_j$ of $j$-point vertices is given.  The count is also used to obtain 
the number of classical graphs in gauge theory and gravity.

\vfill\break

A count of scalar field theory graphs in a an arbitrary scalar Lagrangian is 
exactly given.  The count also produces the relevant number of tree graphs 
in gauge theory and general relativity.  Some well known work in this 
direction is contained in the articles \cite{BenderWu}.  

The count is produced by coupling a scalar field theory to an auxiliary field 
that couples vertices together through the internal lines of the tree 
diagrams.  The number of $m$-point vertices is deduced by keeping track of 
the coupling constant of the scalar and auxiliary field interaction. 

The initial Lagrangian is, 

\bqr 
{\cal L} = {1\over 2} \xi^2 + \sum_{j=3}^M \sum_{a=1}^{j-1} 
 \lambda_j \xi^a \phi^{j-a} {1\over a!(j-a)!} \ , 
\label{initialL} 
\fqr 
with the interactions $\phi^n \xi^m$.  The $\lambda_j$ couplings enter the 
graphs, with external $\phi$ lines and internal $\xi$ lines.  Integrating 
out the auxiliary $\xi$ field produces all of these classical contributions.  

The final Lagrangian is a function ${\cal L}(\phi,\lambda_j)$; expanding 
it in terms of the scalar field and the couplings $\lambda_j$ generates 

\bqr 
{\cal L}_f = \sum  a_{n,\{n_j\}} \phi^n \prod_{j=1}^\infty \lambda_j^{n_j} \ , 
\label{expandedL}
\fqr 
and the number $n_j$ tells how many $j$-point vertices there are in the 
diagram with $n$ external lines.  An $n$-point graph with $\sum n_j$ vertices 
has the multiplicity of $a_{n,\{n_j\}} n!$; this number is derived by 
integrating out the $\xi$ field and expanding the expression in $\phi$ and 
the vertex couplings.

\begin{figure}
\begin{center}
\epsfxsize=12cm
\epsfysize=8cm
\epsfbox{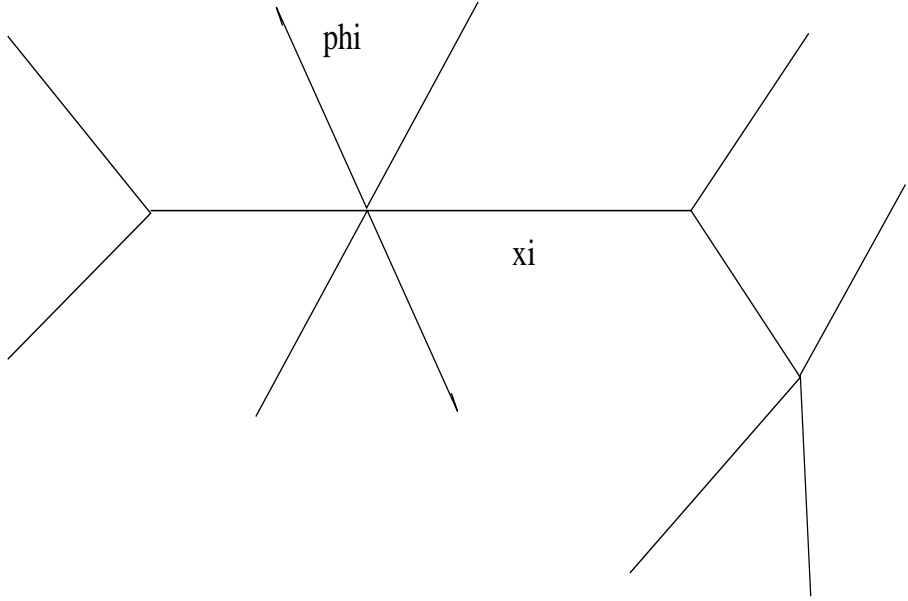}
\end{center}
\caption{Example diagram with internal auxiliary field $\xi$ and external 
scalar $\phi$ fields.} 
\end{figure}

The equations of motion of $\xi$ after a shift of $\xi\rightarrow \xi-\phi/2$ 
is, 

\bqr 
0 = \xi-\phi/2 + \sum_{j=3}^M {\lambda_j\over (j-1)!} 
  \Bigl( (\xi+\phi/2)^{j-1} - (\xi-\phi/2)^{j-1} \Bigr) \ . 
\label{equationmotion}
\fqr 
In the limit of large vertex number $M$, i.e. $M\rightarrow\infty$, and when 
the couplings are identified $\lambda_j=\lambda$, the form in 
\rf{equationmotion} is, 

\bqr 
0 = \xi-\phi/2 + \lambda\Bigl(-\xi-\phi/2 + \xi-\phi/2\Bigr) + 
  \lambda\Bigl(e^{\xi+\phi/2}-e^{\xi-\phi/2}\Bigr) \ , 
\fqr 
or, 
\bqr 
0 = \xi + (-1/2-\lambda) \phi + 2 \lambda e^\xi \sinh(\phi/2)  \ . 
\label{summedeqmotion}
\fqr 
Shifting back $\xi\rightarrow \xi+\phi/2$ generates, 

\bqr 
0 = \xi - \lambda \phi + 2 \lambda e^\xi e^{\phi/2} \sinh(\phi/2) \ . 
\label{transeqmotion}
\fqr 
This is transcendental equation in $\xi$ which has a formal power seriers 
solution.  

The solution to \rf{transeqmotion} is obtained with the use of the Lambert 
solution.  As only the perturbative solution in $\xi$ is required the branch 
cuts are ignored.  Label the parameters as, 

\bqr 
a=1 \quad b=2 \lambda e^{\phi/2} \sinh(\phi/2)  
 \quad c=1 \quad d=-\lambda \phi \ . 
\label{parameters} 
\fqr 
The general solution to this equation is, 

\bqr
\xi = -{1\over ac} \bigl(a {\rm WL}({cb\over a} e^{cd/a})+cd\bigr) \ ,
\label{solution} 
\fqr 
which in the case of interest in \rf{transeqmotion} is, 

\bqr 
  \xi = - {\rm WL}(b e^d)+d  \ . 
\fqr 
The Lambert function $WL(x)$ has the power series expansion of 

\bqr 
W(x)=\sum_{n=1}^\infty (-1)^{n-1} {n^{n-2}\over (n-1)!} x^n  \ ,  
\fqr 
and leads to the form, 

\bqr 
 \xi = d - \sum_{n=1}^\infty (-1)^{n-1} {n^{n-2}\over (n-1)!} b^n e^{nd}
\fqr 
\bqr 
 \xi = -\lambda \phi - \sum_{n=1}^\infty (-1)^{n-1} {n^{n-2}\over (n-1)!} 
 2^n \lambda^n \sinh^n(\phi/2) e^{n\phi(1/2-\lambda)} \ .  
\label{xisolution} 
\fqr 
The solution to $\xi$ is used to resubstitute into the initial Lagrangian to 
obtain the power series expansion in $\phi$ and $\lambda$.   

The expansion in $\phi$ and $\lambda$ is 
\bqr 
{\cal L} = \sum  a_{n,m} \phi^n \lambda^m \ ,  
\fqr 
and counts the diagrams with multiple vertices at $n-2$-point.  The count 
of diagrams can be obtained with any arbitrary vertex configuration, i.e. 
any numbers $n_j$, with the $\lambda$ coupling only; that is, this is 
simpler than retaining all of the $\lambda_j$ couplings.  An $n$-point 
diagram with only $3$-point vertices has $n_3=n-2$ and the $n$-point 
diagram with only a $3$-point and $n-2$ vertex has $n_3=n_{n-2}=1$.  These 
are the vertex bounds without external $\xi$ fields.  Define $m=\sum n_j$ 
as the sum of all of the $n_j$ vertices; then $a_{n,m}$ counts the diagrams 
for $m$ at $n$ point.  A normalization factor of $n!$ is required to 
compensate for the symmetry factor in the final scalar Lagrangian. 

As an example, integrating out the auxiliary field that contributes to 
four-point diagrams gives, 

\bqr 
{1\over 2} \xi^2 + {1\over 2} \xi \phi^2 \rightarrow -{1\over 2^3} \phi^4 \ .  
\fqr 
The symmetry factor of $3$ comes from the three diagrams that contribute, 
those in the $s$, $u$, and $t$ channels. Thus, $4!$ times $2^{-3}$ equals 
$3$, and this is how many diagrams there are at four-point which stem from 
the three-point vertices.  The more general case is examined with the use 
of solving the transcendental equation and re-inserting the solution into 
the Lagrangian in \rf{initialL}.  

The series expansion complicated by the multiple sums involved.  A Taylor 
series expansion is used to extract the appropriate $n$ and $m$.   

\bqr 
{\cal L}_f^{(1)}= {1\over 2} \Bigl(\lambda \phi + \sum_{n=1}^\infty (-1)^{n-1} 
  {n^{n-2}\over (n-1)!} 2^n \lambda^n \sinh^n(\phi/2) e^{n\phi(1/2-\lambda)} 
   \Bigr)^2
\fqr 
\bqr \hskip -.2in
{\cal L}_f^{(2)} = \lambda \sum_{j=3}^\infty \sum_{a=1}^{j-1}(-1)^a 
    \phi^{j-a} {1\over a!(j-a)!}  \sum_{b=0}^a {a!\over b!(a-b)!}
  \lambda^{b-a} \phi^{b-a} 
\fqr 
\bqr  
\times \Bigl( \sum_{n=1}^\infty (-1)^{n-1} {n^{n-2}\over (n-1)!}
 2^n \lambda^n \sinh^n(\phi/2) e^{n\phi(1/2-\lambda)}\Bigr)^b                  
\label{infiniteterms}
\fqr 
The counting numbers $a_{n,m}$ are then derived from 

\bqr 
a_{n,m} = {1\over n!m!}~ \partial_\phi^n \partial_\lambda^m 
 \Bigl( {\cal L}_f^{(1)} + {\cal L}_f^{(2)}\Bigr) \ .  
\fqr 
These derivatives are straightforward to find, but tedious. 

The $\partial_\phi^n$ derivatives distribute in the manner, 
with $n=\alpha_1+\sum^{\alpha_2} \beta_i$,   

\bqr 
H^{\alpha_1,\alpha_2;p_i;\beta_i}=\partial_\phi^{\alpha_1} \phi^{b+j-2a} 
\prod_{i=1}^{\alpha_2} 
 \partial_\phi^{\beta_i} 
   \Bigl(\sinh^{p_i}(\phi/2) e^{p_i\phi(1/2-\lambda)}\Bigr)  
\fqr 
\bqr 
= {(b+j-2a)!\over (b+j-2a-\alpha_1)!} \delta_{b+j-2a,\alpha_1} 
 \times \prod_{i=1}^{\alpha_2} 
 2^{-p_i} \sum_{q=0}^{p_i} {p_i!\over q!(p_i-q)!} (-1)^{p_i-q} 
   (q-p_i\lambda)^{\beta_i} 
\label{individualterms} \ , 
\fqr 
with the $b$th power of the sum in the parenthesis changing to 
$b-\alpha_2$ and a factor present of $b!/(b-\alpha_2)!$.  
All possible combinations are required, including the $\alpha_2$ 
factors from the sums of the object in the parenthesis.  This results, 
after evaluating the expression at $\phi=0$, in 

\bqr 
 \lambda \sum_{j=3}^\infty \sum_{a=1}^{j-1}(-1)^a 
   {1\over a!(j-a)!}  \sum_{b=0}^a {a!\over b!(a-b)!} \lambda^{b-a}
\fqr 
\bqr  
\times 
\sum_{\alpha_1,\alpha_2;\beta_i}^{n=\alpha_1+\sum \beta_i}
\delta_{b,\alpha_2} {b!\over (b-\alpha_2)!} ~
~\sum_{\tilde p=p_1+\ldots+ p_{\alpha_2}; p_i} ~ 
 (-1)^{{\tilde p}+\alpha_2}\lambda^{\tilde p}
  \prod_{i=1}^{\alpha_2} {{p_i^{p_i-2}}\over (p_i-1)!} 
\fqr 
\bqr 
\times H^{\alpha_1,\alpha_2;p_i;\beta_i}
\label{phiderivatives}
\fqr 
The sums include $n=\alpha_1+\sum^{\alpha_2} \beta_i$, and $\alpha_2$ 
ranging from $1$ to $n$.  The $\beta_i$ must each be at least $1$.  
There appear to be indefinite sums on the $p_i$ due to the nested sums in 
the parenthesis of \rf{infiniteterms}.

The coupling derivatives act as 

\bqr 
\partial_\lambda^m ~\lambda^{b-a+{\tilde p}} \prod_{i=1}^{\alpha_2} 
  (q-p\lambda)^{\beta_i}  
\fqr 
\bqr 
 = \sum_{r=0}^m {(b-a+{\tilde p})!\over (b-a+{\tilde p}-r)!} 
 \delta_{b-a+{\tilde p},m}
~\sum_{\gamma_i}  
  \prod_{i=1}^{\alpha_2} {\beta_i!\over (\beta_i-\gamma_i)!} 
 q^{\beta_i-\gamma_i} (-p)^{\gamma_i} 
 \vert_{\sum \gamma_i=b-a+{\tilde p}-r} \ .  
\label{lambdaderivatives}
\fqr 
These derivatives are taken on the couplings in \rf{phiderivatives}, 
and the factors must be placed together.  The two results, in 
\rf{phiderivatives} and \rf{lambdaderivatives}, generate the expression 
for $a_{m,n}$ after dividing by $m!n!$.

The total contribution to $a_{n,m}$ from ${\cal L}_2$ is the complicated 
expression,

\bqr 
 \lambda \sum_{j=3}^\infty \sum_{a=1}^{j-1}(-1)^a 
   {1\over a!(j-a)!}  {a!\over \alpha_2!(a-\alpha_2)!}
\fqr 
\bqr  
\times 
\sum_{\alpha_1,\alpha_2;\beta_i}^{n=\alpha_1+\sum \beta_i}~
~\sum_{\tilde p=m+a-\alpha_2; p_i} ~ 
 (-1)^{{\tilde p}+\alpha_2}
  \prod_{i=1}^{\alpha_2} {{p_i^{p_i-2}}\over (p_i-1)!} 
\fqr 
\bqr 
{(\alpha_2+j-2a)!\over (\alpha_2+j-2a-\alpha_1)!} 
 \delta_{\alpha_2+j-2a,\alpha_1} 
 \times \prod_{i=1}^{\alpha_2} 
 2^{-p_i} \sum_{q=0}^{p_i} {p_i!\over q!(p_i-q)!} (-1)^{p_i-q} 
\fqr 
\bqr 
  \sum_{r=0}^m {m!\over (m-r)!} 
~\sum_{\gamma_i}  
  \prod_{i=1}^{\alpha_2} {\beta_i!\over (\beta_i-\gamma_i)!} 
 q^{\beta_i-\gamma_i} (-p)^{\gamma_i} 
 \vert_{\sum \gamma_i=b-a+{\tilde p}-r} \ .  
\fqr 
The contribution from ${\cal L}_1$ is similar and can be determined 
from the same operations.   $\alpha_1$ and $\alpha_2$ are defined as 
before.  The sums also could be simplified more, and the large 
$n$ expansion generates the known exponential dependence.  

These combinatorical factors can also be used to find the multiplicity 
of gauge 
and gravity diagrams.  In Gervais-Neveu guage, the 3-point and 4-point 
vertices of a non-abelian gauge theory contain $6$ and $12$ terms in a 
diagram without color ordering.  The individual scalar field contributions 
with $n_3$ and $n_4$ vertices ($3$ and $4$-point) expand into an additional 
$3^{n_3} 4^{n_4}$ combinations; the propagator contains only a single 
$\eta_{\mu\nu}$ and does not cause further multiplicity.

Closed form expressions for graph multiplicities in quantum field theory 
are given.  The count pertains to scalar field theory with any number 
and type of vertices.  The number of diagrams $a_{n,m}$ counts the classical 
field diagrams at $n$-point containing $\sum v_i=m$ vertices, with $i$ 
being the number of lines at each vertex.  The count is useful for determining 
the naive complexity of tree graph calculations in quantum field theory.  
Also, the count is useful in determining K\"ahler potentials in toric 
varieties, which requires the number of tree graphs at zero momentum in 
order to determine the D-terms.  

\vfill\break


\begin{thebibliography}{99}

\bibitem{BenderWu}  
C.~M.~Bender and T.~T.~Wuk, Statistical Analysis Of Feynman Diagrams, 
Phys.\ Rev.\ Lett.\  {\bf 37}, 117 (1976); Large Order Behavior Of 
Perturbation Theory, Phys.\ Rev.\ Lett.\  {\bf 27}, 461 (1971).

\bibitem{Chalmers1} 
G. Chalmers, {\it Quantum Solution to Scalar Field Models}, physics/0505018.  

\bibitem{Chalmers2} 
G. Chalmers, {\it Quantum Gauge Amplitude Solutions}, physics/0505077.  

\bibitem{Chalmers3} 
G. Chalmers, {\it Tree Amplitudes in Gauge and Gravity Theories}, 
physics/0504219.  

\bibitem{Chalmers4} 
G. Chalmers, {\it Tree Amplitudes in Scalar Field Theories}, physics/0504173.  

\bibitem{Chalmers5}
G. Chalmers, {\it Derivation of Quantum Field Dynamics}, physics/0503062.  


\end{thebibliography}
\end{document}